\documentclass[runningheads]{llncs}
\usepackage[T1]{fontenc}
\usepackage{amsmath,amssymb,amsfonts}
\usepackage{algorithmic}
\usepackage{csquotes}
\usepackage{calc}
\usepackage{ifthen}
\usepackage{xspace}
\usepackage{graphicx}
\usepackage{textcomp}
\usepackage{xcolor}
\usepackage[inline]{enumitem}
\usepackage{siunitx}
\usepackage{hyperref}
\usepackage[normalem]{ulem}
\usepackage[font={footnotesize}]{caption}
\def\BibTeX{{\rm B\kern-.05em{\sc i\kern-.025em b}\kern-.08em
  T\kern-.1667em\lower.7ex\hbox{E}\kern-.125emX}}

\newcommand{\ATA}{\mysize{\text{A\!T\!A}}\xspace}
\newcommand{\ATAs}{\mysize{\text{A\!T\!As}}\xspace}
\newcommand{\scale}{.95}

\newcommand{\mysizetxt}[1]{\scalebox{\scale}{#1}}
\newcommand{\mysize}[1]{\scalebox{\scale}{\ensuremath{#1}}}

\newcommand{\mykey}[1]{\texttt{\mysizetxt{#1}}}

\newcommand{\myttt}[1]{\mysize{\texttt{#1}}}

\newcommand{\agentA}{\textit{A}\xspace}
\newcommand{\agentE}{\textit{E}\xspace}

\newcommand{\goals}{\texttt{\mysizetxt{goals}}}
\newcommand{\beliefs}{\texttt{\mysizetxt{beliefs}}}
\newcommand{\behaviour}{\texttt{\mysizetxt{behaviours}}}

\newcommand{\MODD}{\mysizetxt{VODD}\xspace}
\newcommand{\MODDs}{\mysizetxt{VODDs}\xspace}

\newcommand{\other}{\texttt{\mysizetxt{other}}\xspace}
\newcommand{\expl}[1]{\emph{#1}}

\newcommand{\goalE}{\ensuremath{\mykey{g}_t}\xspace}
\newcommand{\goalS}{\ensuremath{\mykey{g}_c}\xspace}

\newcommand{\Aware}{\footnotesize\texttt{(aware)}\normalsize}
\newcommand{\Transfer}{\footnotesize{\texttt{(transfer)}}\normalsize}
\newcommand{\Resolve}{\footnotesize{\texttt{(resolve)}}\normalsize}

\newcommand{\ie}{i.e.\xspace}

\newcommand{\eg}{e.g.\xspace}

\newcommand{\cf}{cf.\xspace}

\usepackage{ifthen,comment,amssymb}
\usepackage{enumitem}
\usepackage[british]{babel} 
\usepackage{floatrow}
 
\begin{document}

\title{Designing Value-Aligned Traffic Agents through Conflict Sensitivity}
 	

\author{Astrid Rakow\inst{1}  \and
Joe Collenette\inst{2}  \and
Maike Schwammberger\inst{3}\and
Marija Slavkovik\inst{4}  \and
Gleifer Vaz Alves\inst{5} 
}

\authorrunning{A. Rakow et al.}
%
%
\institute{ German Aerospace Center, Institute of Systems Engineering for Future Mobility, Oldenburg, Germany, \email{astrid.rakow@dlr.de} \and
School of Computer and Engineering Sciences, University of Chester, United Kingdom, \email{j.collenette@checster.ac.uk} \and
Karlsruhe Institute of Technology (KIT), Germany, \email{schwammberger@kit.edu} \and
University of Bergen, Norway, \email{marija.slavkovik@uib.no} \and
Universidade Tecnologica Federal do Paran\'{a}, Brasil, \email{gleifer@utfpr.edu.br}}
\maketitle      
\begin{abstract}
	Autonomous traffic agents (\ATAs) are expected to act in ways that are not only safe, but also aligned with stakeholder values across legal, social, and moral dimensions. In this paper, we adopt an established formal model of conflict from epistemic game theory to support the development of such agents. We focus on value conflicts--situations in which agents face competing goals rooted in value-laden situations and show how conflict analysis can inform key phases of the design process. This includes value elicitation, capability specification, explanation, and adaptive system refinement. We elaborate and apply the concept of \emph{Value-Aligned Operational Design Domains} (\MODDs) to structure autonomy in accordance with contextual value priorities. Our approach shifts the emphasis from solving moral dilemmas at runtime to anticipating and structuring value-sensitive behaviour during development.
\end{abstract}


                                             
\section{Introduction}\label{sec:intro}
Autonomous traffic agents (ATAs)--automated systems with high levels of auto\-nomy operating in traffic environments--are increasingly expected to behave not only safely, but also aligned with evolving values, such as legal rules, safety, social conventions, and moral principles.
Rather than treating value alignment as an afterthought, we advocate for treating \emph{conflicts}--situations where no available option satisfies all relevant goals or values--as a central concept in the design of autonomous systems.

Building on the formal conflict model from ~\cite{conflict}, our approach operationalises values as design-time goals that agents are expected to pursue. This allows value alignment to be analysed within a goal-based conflict framework, enabling the detection and resolution of tensions between competing values during both development and runtime


Focusing on \emph{value conflicts}, in which not all value-induced goals can be jointly satisfied, enables developers to identify, explain, and constrain misalignments across the system lifecycle. We refer to an agents capacity to recognise, reason about, and act on such tensions as \emph{conflict sensitivity}, a crucial aspect of realising value-aligned behaviour.

To realise structured value-aligned autonomous behaviour, we introduce \emph{Value-Aligned Operational Design Domains} (\MODDs). Traditional ODDs define operational boundaries but fail to account for value-laden trade-offs--such as those between efficiency and fairness. \MODDs address this by incorporating not only operational feasibility but also normative expectations, structuring autonomy around prioritised goals and specifying when autonomous value-aligned decision-making or control transfer is required.

By treating value conflicts as a central design concern, our approach supports structured reasoning about agent capabilities and autonomy boundaries. The conflict model provides a foundation for assisting agents in making value-aligned decisions and delegating control in a context-aware manner.

\vspace{0.5em}
\noindent
This paper makes three main contributions:
First, we demonstrate how a formal conflict model can support the development of value-aligned behaviour in \ATAs;
Second, we introduce the concept of \emph{Value-Aligned Operational Design Domains} (\MODDs), which represent scopes of autonomous decision-making guided by goal hierarchies and contextual conditions for transferring control, if required;
Third, we propose an architectural vision for conflict-sensitive \ATAs and identify key challenges for enabling agents to detect, explain, and act upon conflicts throughout their lifecycle.

\paragraph{Outline}
\autoref{sec:soa} reviews related work on conflict models and value alignment in autonomous systems, highlighting gaps in supporting morally and socially aware behaviour in open-world scenarios.
\autoref{sec:conflict} introduces the formal conflict notion adopted in our work and relates it to value-alignment in socially complex settings, including those involving moral reasoning.
\autoref{sec:development} outlines a structured approach to designing socially acceptable agents based on conflict sensitivity.
\autoref{sec:FromConf} formalises Value-Aligned Operational Design Domains (\MODDs). 
\autoref{sec:challenges} identifying open challenges and future directions. \autoref{sec:concl} concludes with a summary.

\section{Related Work}\label{sec:soa}
This section surveys work on conflict models and value alignment in autonomous traffic agents (\ATAs). While many approaches address physical or planning-level conflicts, fewer focus on resolving value tensions in morally and socially complex environments. We review how the notion of conflict is used across related work, how agents reason in open contexts, and how value alignment is supported--or neglected--in current design frameworks.

\paragraph{Conflicts in Traffic Domains.}
Initially, traffic conflicts were defined in terms of potential collisions~\cite{TrafficConflict,Chin1997,Glauz1980}. Early work addressed path planning and collision avoidance~\cite{Hwang.1992,Raja.2012}, later expanding to coordination among traffic agents~\cite{RiosTorres.2017} and airspace scenarios~\cite{Albaker.2010}. Conflict-Based Search (CBS)~\cite{Sharon.2012,Li.2019,Kottinger.2022} became central in multi-agent pathfinding, where conflicts guide route optimisation. With autonomous vehicles, the notion of conflict came to include ethical~\cite{Bonnefon.2016,Lin.2015,Goodall.2014}, legal~\cite{Gurney.2013,Koopman.2017}, and social~\cite{Lee.2004} considerations.

\paragraph{Formal Conflict Notions.}
Our approach builds on the formal conflict model by Damm et al.~\cite{conflict}, which applies Galtung's conflict triangle~\cite{Galtung1969} in an epistemic game-theoretic framework. Conflicts are characterised through incompatible goals, beliefs, and actions. How this model can support explanation generation to mitigate risk is shown in ~\cite{Explain2025}.

Other approaches include Pawlak's conflict theory ~\cite{pawlak}, Deja's attitude-based extension~\cite{Deja2000}, and Zurek and Wyner's value-oriented argumentation ~\cite{Zurek22}. Evans et a.~\cite{Evans2020} model moral dilemmas in autonomous driving by balancing harm and uncertainty across multiple ethical profiles.

\paragraph{Ethics in Open Contexts.}
Autonomous agents operating under open-world assumptions must make decisions that go beyond pre-programmed rules. The SocialCars project~\cite{SocialCars} emphasized decentralised, cooperative behaviour to address social dynamics in traffic systems.
Value Sensitive Design (VSD)~\cite{VSD2006} offers a  theory and method framework that considers human values throughout the design process. It integrates stakeholder driven conceptual, empirical, and technical investigations to ensure that the resulting system aligns with diverse value concerns. While VSD explicitly recognises and seeks to balance value tensions, it lacks a formal treatment of dynamic or adversarial value conflicts. Standards such as ISO/IEC/IEEE 24748-7000~\cite{P7000} highlight the need for more operational mechanisms in ethical system design.
Our work complements VSD by introducing a formal conflict model that operationalises values as goals constraining actions based on beliefs. This enables detection, explanation, and mitigation of value tensions throughout the agent lifecycle.

\paragraph{Operational and Value-Aligned Design Domains.}
Operational Design Domains (ODDs), as defined in SAE J3016~\cite{sae_j3016} and formalised in ISO 34503~\cite{iso34503}, specify the operational boundaries--environmental, temporal, and functiona--within which autonomous systems are considered to be safe to operate.
ISO 21448 (SOTIF)~\cite{ISO21448.2022} extends this safety view by accounting for hazards resulting from intended functionality under uncertain or incomplete knowledge.

We introduce \MODDs (\cf \autoref{sec:FromConf}) as a complementary abstraction to ODDs.
They support socially acceptable autonomy control in value-sensitive and uncertain contexts.
\MODDs define when a value hierarchy should guide autonomous decisions and when control must be deferred.
Within an ODD, they support value-aligned decision-making when multiple functionally viable options exist, and at its boundary, they provide autonomy control for unforeseen situations.
This dual function helps to align technical and normative considerations.

The concept of constraining the extend of moral autonomy through contextual boundaries was introduced as Moral ODDs in earlier work~\cite{moralTraffic}, though only in preliminary form. That work outlined the idea conceptually but did not formalise its integration into design or decision-making. Here, we extend and operationalise the notion through explicit value hierarchies and behavioural policies.
\paragraph{Conclusion}
While prior work has explored traffic conflicts, formal models, and value-sensitive design, many approaches do not provide systematic support for design and autonomy adaptation in open contexts. Our work addresses this by formalising conflict as a dynamic interplay of \goals, \beliefs, and \behaviour. This enables capturing value tensions across the design, validation, and operational phases.

\section{Conflicts and Moral Reasoning}
\label{sec:conflict}
In this section, we introduce the epistemic game-theoretic conflict model proposed by Damm et al.~\cite{conflict}.
 We illustrate the model's applicability to complex decision-making, and connect it to ethical reasoning. In doing so, we provide a conceptual foundation for informing value-aligned agent design, as developed in the following sections.
 
\begin{figure}[h]
	\floatbox[{\capbeside\thisfloatsetup{capbesideposition={right,top},capbesidewidth=6.8cm}}]{figure}[\FBwidth]
	{\vspace*{2.5mm}\null\caption{\small 
			A conflict situation involving incompatible goals: two vehicles approach a narrowing road, both intending to proceed and avoid collision. Resolution requires agreement on who yields.}\label{fig:opposing}}
	{\includegraphics[width=\textwidth-7.3cm]{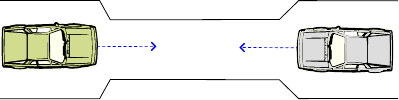}}
\end{figure}

Conflicts in traffic are frequent and vary in severity. Consider the case in \autoref{fig:opposing}: two vehicles approach a narrowing road simultaneously. Each agent aims to \mykey{(g1)} proceed and \mykey{(g2)} avoid collision--yet both cannot achieve these goals without coordination.

More subtle conflicts arise in decisions involving moral or stakeholder values. \expl{For example, an \ATA may face two evasive manoeuvres: one risks hitting a pet, the other damaging the vehicle. Resolving such dilemmas requires recognising and weighing the involved values.} These cases illustrate that many conflicts concern value trade-offs shaped by cultural norms, legal obligations, and situational factors. Unlike traffic rules or utilitarian heuristics, moral values express socially grounded expectations that vary by context. Reflecting these values is important for social acceptability and for providing robust guidance in unforeseen situations but doing so remains a significant design challenge.

Our conflict-sensitive perspective is based on the conflict notion of Damm et al.~\cite{conflict}. They define a conflict as a situation in which an agent experiences difficulty in deciding what to do since, based on its beliefs, no available strategy clearly outperforms alternative strategies in terms of goal satisfaction.
The formalisation in \cite{conflict} instantiates Galtung's classic \emph{Conflict Triangle}~\cite{Galtung1969}m which conceptualises conflict through the interrelated dimensions of incompatible \goals, opposing \behaviour, and inconsistent \beliefs. To this end, each agent in \cite{conflict} is reperesenteed through its actions, prioritised goals, and possibly incorrect beliefs corresponding respectively to Galtung's three dimensions. 

Formally, a conflict arises when agents, given their beliefs, cannot identify a dominant strategy--one that outperforms all alternatives in satisfying the goals. The model represents both \emph{actual} and \emph{believed} conflicts\footnote{In a believed but not actual conflict, a joint solution exists.} by modelling both ground truth and agent beliefs. 
In this paper, we show how this formalisation can support the diagnosis of conflict causes, the simulation of conflict emergence, and the exploration of resolution strategies such as belief disclosure, situational alignment, or goal revision.

To operationalize conflict-aware behaviour throughout an agent's lifecycle, we distinguish three core processes: (i) \emph{conflict resolution}, which refers to the process by which an agent selects or executes an action to resolve an ongoing or anticipated conflict; (ii) \emph{conflict delegation}, which refers to the assignment of responsibility for resolving a conflict to an external authority (e.g., a human driver or ethical oversight component); and (iii) \emph{conflict transfer}, which refers to the operational handover of control to another authority.

As this conflict model captures both actual and believed conflicts, it is particularly well-suited for open contexts like traffic--where agents operate with limited observability and uncertain causal knowledge. In such settings, agents must reason based on beliefs about others' intentions, goals, and values while accounting for uncertainty. Many of these values reflect underlying moral considerations and social norms. To develop value-aligned agents, decision-making must therefore incorporate moral reasoning and normative sensitivity.

\subsection{Conflicts Involving Moral Reasoning}\label{sec:moral}

Socially acceptable \ATAs must be sensitive to moral and evolving social norms. Moral reasoning is hence a central challenge for socially acceptable \ATAs.
In this subsection, we abstract from implementation details to discuss conflicts in terms of general agents and their moral reasoning capabilities. The implications for \ATAs are addressed in the subsequent sections.

Successful operation in traffic requires the ability not only to resolve conflicts with other traffic participants but also to handle moral dilemmas that may arise \cite{Greene2016,Kirkpatrick2015,Millard-Ball2018}. Making ethical choices can be understood as taking into account the interests of others in addition to one's own goals, preferences, and constraints.
Moral conflicts or moral dilemmas are typically defined as situations, in which all of the available choices to the agent, including not doing anything, are
in some way not ethical~\cite{Yu18,Liao23}. The moral dilemmas either do not have a solution
or the agent needs to reason, identify and justify doing the least unethical thing~\cite{DennisFSW16}.

While classical moral dilemmas are rare in everyday traffic, conflicts involving trade-offs among normative values are common. \emph{Consider a truck overtaking another on the motorway: the slower truck may reduce its speed to support smoother traffic flow, even though its goal is timely delivery and no regulation requires it.} Such decisions illustrate how the enactment of (traffic) rules can be shaped by social and moral values.

We use the term \emph{value conflict} to refer to situations in which all available options compromise some relevant values. Such conflicts may involve interpersonal disagreements over values or intra-agent tensions between multiple values an agent is expected to uphold. The values may include moral, legal, safety, and social concerns. Thus, value conflicts encompass classical ethical dilemmas, but also more routine trade-offs, where the agent needs to weigh and prioritise context-dependent normative expectations.

A value-aligned agent \agentA should uphold relevant values--its own and those it adopts from its social environment. To do so, the agent must act on goals derived from those values.
To capture value conflicts, we consider a special agent \agentE that embodies \agentA's internalised moral and social obligations. \agentE's goals reflect values that may not be externally enforced but should still influence \agentA's behaviour.
Thus, a value conflict becomes a goal conflict between agent \agentA and agent \agentE. But unlike a classical agent, \agentE cannot hinder \agentA from pursuing its goals, \ie, disregarding \agentE is always an option.
The core challenge is how to characterise the resolution of such conflicts. A value-aligned agent must act on normative goals without being compelled to prioritise them in every decision.
\emph{What does it mean, precisely, to uphold values? How often, and under what circumstances, may the goals of \agentE be disregarded before the agent \agentA is seen as misaligned?}

One practical approach to approximate value alignment is to define \emph{value-informed goal patterns} that specify how frequently or under which conditions certain value-driven goals must be achieved. \emph{For example, one might define a pattern in which environmental concerns are prioritised over driving efficiency for more than half of the total distance travelled. }

\begin{figure}
	\includegraphics[width=\textwidth]{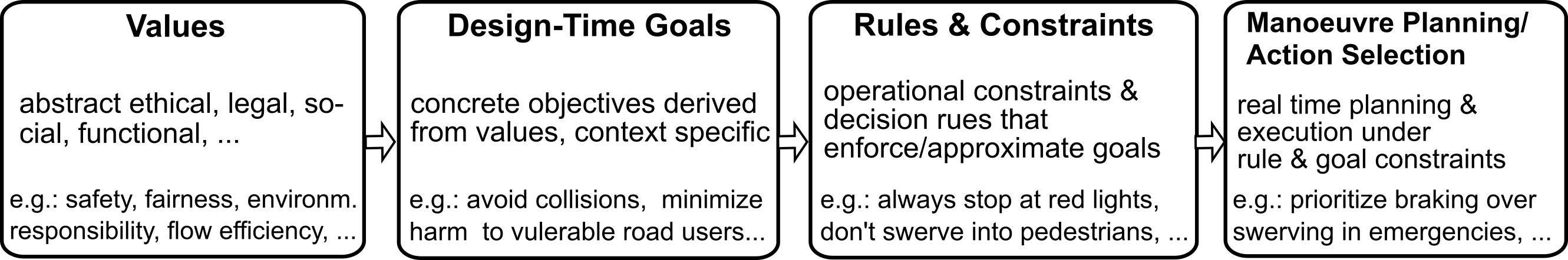}
	\caption{Structuring agent behaviour through value-guided goals and rules.}\label{fig:values2control}
\end{figure}

Although a precise characterisation of value alignment remains open, modelling value conflicts as goal incompatibilities--particularly between agents and non-compulsory normative authorities like \agentE--leads to making them explicit in design. Identifying such conflicts enables developers to scope autonomy and anticipate when agent behaviour might endanger important values, making conflict awareness a practical tool for aligning their behaviour with social expectations. 
To make this approach explicit, Figure~\ref{fig:values2control} illustrates how values are systematically incorporated into agent behaviour. Values are first expressed as prioritised goals, which are then operationalised through rules and constraints--for example, by adjusting system capabilities, defining functional boundaries, or encoding traffic laws and ethical priorities during agent design. The rules and goal constraints constrain the action selection of the agent's real time planning. 

\section{Designing Conflict-Sensitive Agents for Value-Aligned Autonomy}\label{sec:development}
This section shows how the formal conflict model can support key activities in the development of autonomous agents: identifying stakeholder values, designing necessary capabilities, verifying and validating behaviour, and adapting agent behaviour over time.

These activities are supported by conflict scenarios, which expose situations where value-aligned decisions are required. Conflict scenarios are abstract in nature but can be refined into concrete system runs. They can be simulated and visualised \eg through animation or interactive play-outs of Traffic Sequence Charts (TSCs)~\cite{TSCs,scenarioCov,Consistency}, a formal yet visually intuitive specification language to elicit stakeholder feedback. Such scenarios help uncover value tensions, assess trade-offs, and address stakeholder concerns early in the design process. They also support the definition of targeted test cases and provide adaptation triggers after deployment.

\subsection{Value Elicitation}\label{sec:aware}

Value elicitation is a crucial step in the development of ethically aligned systems, as emphasised in standards such as ISO/IEC/IEEE 24748-7000~\cite{IEEE7000} (originating from IEEE P7000~\cite{P7000}) and academic frameworks like Value Sensitive Design (VSD)~\cite{VSD2006}.Value elicitation aims to identify values relevant to stakeholders that should be reflected in system behaviour. It typically involves engaging stakeholders to understand their expectations, preferences, and concerns, and translating these into actionable system requirements.

Autonomous traffic agents (\ATAs) especially lack awareness of the values that human drivers intuitively express in conventional driving. Human behaviour often aligns, consciously or not, with social, personal, and moral values (cf.~\cite{Evans2020} or \autoref{sec:moral}). To ensure that \ATAs behave in ways that reflect societal expectations, these values must be made explicit through value elicitation. As Evans et al.~\cite{Evans2020} note, values are context-dependent and frequently in conflict, which makes their formalisation particularly challenging.

The following subsections present two complementary approaches to eliciting stakeholder values: one through structured conflict modelling at design time, and one through runtime feedback mechanisms that detect value misalignments.

\subsubsection{Conflict-Driven Design-Time Elicitation}

Rather than relying solely on abstract interviews or predefined value lists, we propose a model-based elicitation approach grounded in formal conflict analysis.

\mykey{(Step 1: Coarse Initial Value Assumptions)} The process begins with a deliberately conservative set of stakeholder value assumptions. These initial values act as provisional bounds, helping to identify situations where the agent's planned behaviour may lead to value tensions.

\mykey{(Step 2: Interaction Modeling)} Developers construct a simplified interaction model targeting typical interactions between the \ATA and stakeholder-representing agents in its environment. Each agent's goals, beliefs, and actions are explicitly encoded reflecting the three dimensions of conflict identified in Galtung's conflict triangle~\cite{Galtung1969}.

\mykey{(Step 3: Conflict Scenario Generation)} Simulation or reachability analysis\footnote{Reachability analysis systematically explores the space of possible system states to determine whether specific conditions such as conflicts can occur.} is used to identify reachable conflict situations, either from the \ATA's perspective or based on the ground truth system state. These scenarios highlight potential value tensions for further analysis. 

\mykey{(Step 4: Scenario Analysis and Stakeholder Engagement)} As the models and value assumptions are only coarse approximations, the generated conflict scenarios reflect potential, but not guaranteed, misalignments. These scenarios are then analysed and discussed with stakeholders, either in abstract form or as concrete instances via simulations or interactive play-outs.

\mykey{(Step 5: Iterative Refinement)} Based on stakeholder feedback, both the contextual value profiles and the interaction model are refined. Steps 3 and 4 are iterated until the captured values and interactions sufficiently reflect stakeholder expectations.

Given the complexity of traffic environments, the underlying models must remain abstract yet expressive. We recommend using a Systems-Theoretic Accident Model and Processes (STAMP)-inspired approach~\cite{Stpa2018,Zhang2022} to construct hierarchical representations of responsibilities and interactions. This supports a top-down analysis of the socio-technical context in which the \ATA operates.

\expl{For example, cyclists may feel intimidated by \ATAs that pass \enquote{too closely}. To elicit the values underlying this discomfort, developers model typical cyclist vehicle interactions, including spatial dynamics and the \ATA's beliefs about safe distances. To surface conflicts, the \ATA can initially be modelled as only weakly upholding comfort-related values. Reachability analysis then reveals situations where legally compliant behaviour, driven by these beliefs, still causes discomfort. Presenting such scenarios with varying context (e.g., lighting, road surface) enables stakeholder feedback to refine both the interaction model and value assumptions--revealing comfort factors like splash water, noise, or obstruction as relevant.}

\subsubsection{Runtime Detection of Value Misalignment}\label{sec:runtime}
While design-time analysis allows designers to anticipate value tensions, not all relevant values or conflicts can be foreseen. Runtime monitoring complements this by enabling the detection of value misalignments based on operational data and stakeholder feedback.
The core idea is to continuously observe the \ATA and its surrounding context, logging relevant system states and interactions. This allows stakeholders to flag undesired behaviours either retrospectively or in real time. These reports can then be analysed in conjunction with system logs to identify underlying value conflicts.
A key challenge is determining what to monitor. The formal conflict model offers guidance: by identifying which \goals, \beliefs, and {\behaviour} are causally relevant to a conflict, it informs the specification of what information should be logged. However, if the system model lacks critical contextual detail, some conflicts may not be reconstructable post hoc, which shows that runtime feedback can also inform model refinement.

\expl{For example, suppose a cyclist reports feeling intimidated by an \ATA passing too closely. If only basic spatial data (e.g., distance, speed) were logged, the cause may remain unclear. However, if contextual factors like lighting, road conditions, or splashing were recorded or added in response to the report developers can iteratively refine the model until the conflict becomes explainable. This process can surface stakeholder values that were not captured during initial design.}

Such iterative, feedback-driven refinement deepens the understanding of conflicts and supports developers in adjusting the system's autonomy boundaries, behavioural policies, or explanatory capabilities based on runtime evidence.

\subsection{Conflict-Adjusted Capabilities}\label{sec:capas}

Designing conflict-sensitive, value-aligned \ATAs requires not only the identification of relevant values, but also the capabilities required to act in accordance with them. The conflict model--structured around \mykey{(}\goals, \beliefs, \behaviour\mykey{)}--provides levers for design adjustments of the \ATAs' capabilities.

\subsubsection{Dissolving Conflicts by Capability Adjustments}\label{sec:capa_adjust}

When model-based analysis reveals that an \ATA encounters unresolvable conflicts, the agent can be redesigned through targeted adjustments to dissolve the conflict:

\mykey{(beliefs)} If a conflict is only believed \ie, based on incorrect or incomplete beliefs it can often be dissolved by enabling the agent to form more accurate beliefs. This may involve improving perception (e.g., adding sensors), enhancing context interpretation (e.g., refining the world model), or enabling information exchange with other agents.

\mykey{(behaviour)} If the conflict stems from the available action space, capabilities can be extended (e.g., evasive manoeuvres, new signalling options) or constrained (e.g., limiting maximum speed) to avoid high-risk scenarios.

\mykey{(goals)} Conflicts may arise from overly rigid or misaligned design-time goals. Allowing greater flexibility such as permitting detours or relaxing efficiency targets can reduce value misalignment and support better trade-offs.

\expl{For instance, a prototype \ATA \agentA pursues the goals of travel-time efficiency (\goalE) and avoiding discomfort to vulnerable traffic participants (\goalS). A conflict analysis reveals an actual but not believed conflict: close overtakes cause discomfort, yet \agentA remains unaware of the conflict because it cannot detect cyclists discomfort. Designers may respond by: (1) enhancing perception to recognise indicators of discomfort \emph{(\beliefs)}; (2) disabling close overtakes of cyclists \emph{(\behaviour)}; or (3) softening the travel-time goal and prioritising greater overtaking distance in such contexts \emph{(\goals)}. The actual conflict can thus be dissolved by adjusting a single conflict dimension.}

\subsection{Explanation of Conflicts}\label{sec:expl}
Explaining value-sensitive decisions is a key capability of socially acceptable agents, essential for building trust, enabling coordination, and ensuring accountability throughout the system lifecycle.
The formal conflict model can also support the generation of explanations--both during development and at runtime. By structuring conflicts in terms of \goals, \beliefs, and \behaviour, the model enables \ATAs to communicate the reasons behind value-sensitive decisions or unresolved tensions. 

An observed conflict can be explained using the agent's current beliefs and goals:
\expl{\enquote{I attempted manoeuvre $x$ to achieve goal $g$ in situation $s$, but \other did $y$, which prevented me from achieving $x$.}}
This explanation includes \mykey{(goal)} the  goal, \mykey{(belief)} the perceived situation,  and \mykey{(behaviour)} intended behaviour and  the observed conflicting behaviour of the other agent.
For anticipated or predicted conflicts, the agent can reason about divergent beliefs or intentions:
\expl{\enquote{I intend to do $x$ for $g$ in situation $s$, but I believe \other sees the situation as $s'$, pursues $g'$, and will do $y$.}}
In this case, the agent explains the potential misalignment in terms of its assumptions about the other agent's goals, situational beliefs, and likely behaviour.
Such structured explanations clarify why value-sensitive trade-offs were made. They support (i) debugging during development, (ii) communication with external authorities during delegation, and (iii) post-hoc accountability and user trust.


\subsection{From Conflict to Assurance and Adaptation}\label{sec:depl}
We propose to use conflict scenarios as test cases. As in safety-critical domains (cf.~ISO 21448~\cite{ISO21448.2022}), scenario-based testing provides a means to validate acceptable behaviour. Conflict scenarios, derived through model-based analysis or post-deployment feedback, can be used to test whether \ATAs resolve these conflicts appropriately and uphold value priorities under uncertainty.  
Replaying, refining, or abstracting previously encountered conflicts enables agents to adapt their autonomy profiles, behavioural strategies, and value hierarchies over time.

\paragraph{From Design Perspective to Scoping Autonomy.}
The conflict-based design steps outlined above emphasise the complexity and context-dependence of value alignment. To address this, the next section introduces \emph{Value-Aligned Operational Design Domains (\MODDs)} as a means of scoping autonomous behaviour according to situational goal hierarchies derived from value considerations. We present the \MODD concept in detail and discuss capability requirements as well as a general architecture for agents with context-sensitive, dynamic value alignment.

\section{Scoping Autonomy through Value-Aligned Operational Design Domains (VODD)}
\label{sec:FromConf}

In this section we turn to the challenge of dynamically controlling autonomy guided by context-dependent value priorities. We identify two key functions where value alignment plays a central but underexplored role: (i) guiding behaviour in unforeseen contexts, and (ii) supporting value-sensitive choices among functionally admissible options.
To address these, we introduce \emph{Value-Aligned Operational Design Domains} (\MODDs), which extend traditional ODDs by embedding goal hierarchies. These hierarchies are grounded in elicited stakeholder values (\cf~\autoref{sec:aware}) and guide value-sensitive autonomy.

The remainder of this section specifies the structural elements of \MODDs, outlines the capability requirements for agents operating within them (\autoref{sec:req_capa}), and presents an architecture for context-sensitive, value-aligned decision-making (\autoref{sec:architecture}).

\subsection{Defining Value-Aligned Operational Design Domains}\label{sec:MODDs}
Conflicts in traffic often reflect tensions between relevant values. In some situations, especially those involving contested social or normative expectations, autonomous resolution may not be permitted, as recommended by \eg the German Ethics Commission~\cite{ethicsCode}.
To formalise the boundaries of such autonomous decision-making, we introduce \emph{Value-Aligned Operational Design Domains} (\MODDs). \MODDs specify the contextual conditions under which an agent is permitted to act autonomously in accordance with specified value priorities, and when control must be transferred to another authority.
%
Each \MODD specifies:\\[-4mm]

\texttt{(S)} \mykey{Scope of Autonomy}, defined through a (possibly partial) hierarchy of goals, which the \ATA may autonomously sacrifice in line with their relative importance. Goals on the same level are considered incomparable; preferences or context-specific heuristics may later refine their ordering, learning \eg from  user feedback.

\texttt{(D)} \mykey{Domain}, defined through entry\,\myttt{(En)}, invariant\,\myttt{(In)}, and exit\,\myttt{(Ex)} conditions.
Each domain condition is associated with a required confidence level: entry and exit conditions must be met with sufficient confidence before the \ATA can enter or leave the domain, while invariants must be maintained throughout operation. Exit conditions also define the external authority to which control is delegated.  

\begin{example}\label{ex:MODD}
	In the example \MODD in \autoref{fig:MODD}, the \ATA autonomously chooses among reducing emissions, minimising travel time, and supporting smooth traffic flow. The \texttt{(S)}cope of Autonomy specifies that these three goals are initially treated as incomparable in the system's default goal hierarchy, allowing the \ATA to resolve trade-offs based on situational judgement.
	
	To enter the domain, \myttt{(En1)} requires that the \ATA does not plan to leave the highway within 60\textit{seconds}. The following invariant conditions must hold:
	\myttt{(In3)} the \ATA is on the highway in normal traffic,\footnote{We envision formally defined scenarios--e.g., using Traffic Sequence Charts~\cite{TSCs}--to specify conditions like \enquote{normal traffic}.}
	\myttt{(In1)} it does not plan to exit within 30\textit{seconds}, and
	\myttt{(In2)} its mission is to reach location \myttt{Z}.
	
	The driver has set a 20\% tolerance for travel time increase.
	If the \ATA predicts with 85\% confidence that \myttt{(Ex1)} this threshold will be exceeded, it requests driver input.
	If it predicts with 70\% confidence that \myttt{(Ex2)} it must exit the highway within 30\textit{seconds}, it transfers control.
	
	\begin{figure}
		\includegraphics[width=\textwidth]{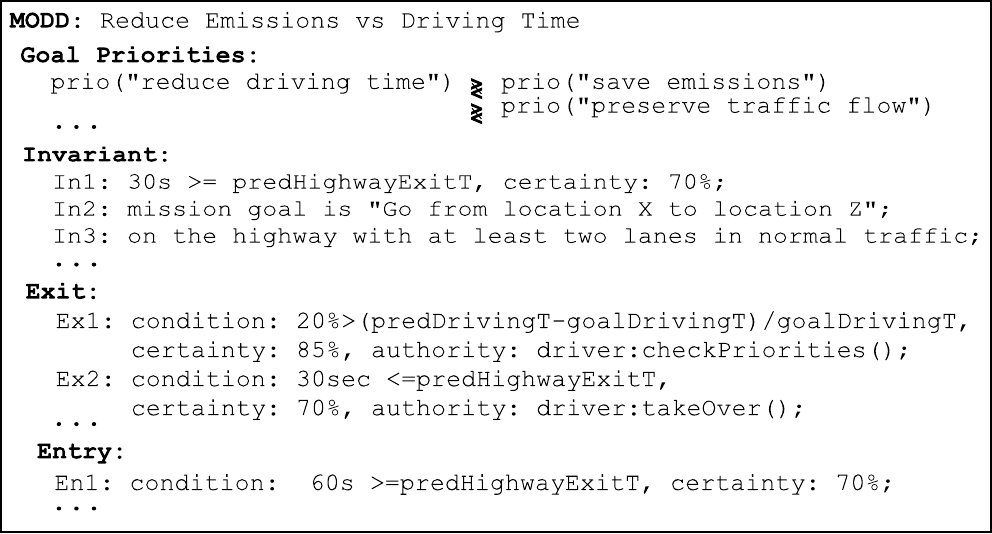}
		\caption{Value-Aligned Operational Design Domain (VODD) with contextual autonomy bounds and autonomy scope allowing trade-offs between emissions reduction, travel time, and support for traffic flow.}
		
		\label{fig:MODD}
	\end{figure}
\end{example}

If a conflict exceeds the scope of the current \MODD, its exit conditions specify how control is delegated--either to another \MODD or to an external authority such as a human driver, remote operator, or oversight system. By linking \MODDs through entry and exit conditions, the system realises adaptive control: both the responsible authority and the applicable value alignment are dynamically determined in response to contextual changes.

\subsection{ATA Capabilities Required for VODD}\label{sec:req_capa} 
To operate reliably within the boundaries defined by \MODDs, an \ATA must exhibit three core capabilities that support conflict-sensitive, value-aligned behaviour:

\begin{enumerate}[labelwidth=!,labelindent*=10mm]
	\item[\Aware\hspace*{5mm}] detect when a conflict arises,
	\item[\Resolve\hspace*{1.5mm}] resolve the conflict if it lies within the current \MODD,
	\item[\Transfer] transfer control if the conflict exceeds the \MODD.
\end{enumerate}
These capabilities require the following functional components:

\emph{Conflict Detection.} The agent must recognise value-relevant aspects of the current context (e.g., splash water, visual obstruction), interpret sensor data, and form beliefs about the situation to detect conflicts.

\emph{Autonomous Resolution.} To resolve conflicts, the agent must enact goal prioritisation as prescribed by the \MODD's hierarchy. 

What such enactment entails in general remains an open question (\cf \autoref{sec:moral}). Approximate encodings--such as value-informed goal patterns-- provide concrete criteria and in some cases allow monitoring, whether the resolutions enact the goal prioritisation.

\emph{Transfer and Explanation.} Conflicts that exceed the \MODD lead to either switching to another \MODD or to the transfer of control to another authority--a driver, remote operator, or ethical module. The transfer must be realized in a safe and timely manner. The conflict must be explained to the recipient for informed decision making.

\subsection{Architecture for Conflict-Sensitive, Value-Scoped Agents}\label{sec:architecture}
\begin{figure}
	\centering
	\includegraphics[width=\textwidth]{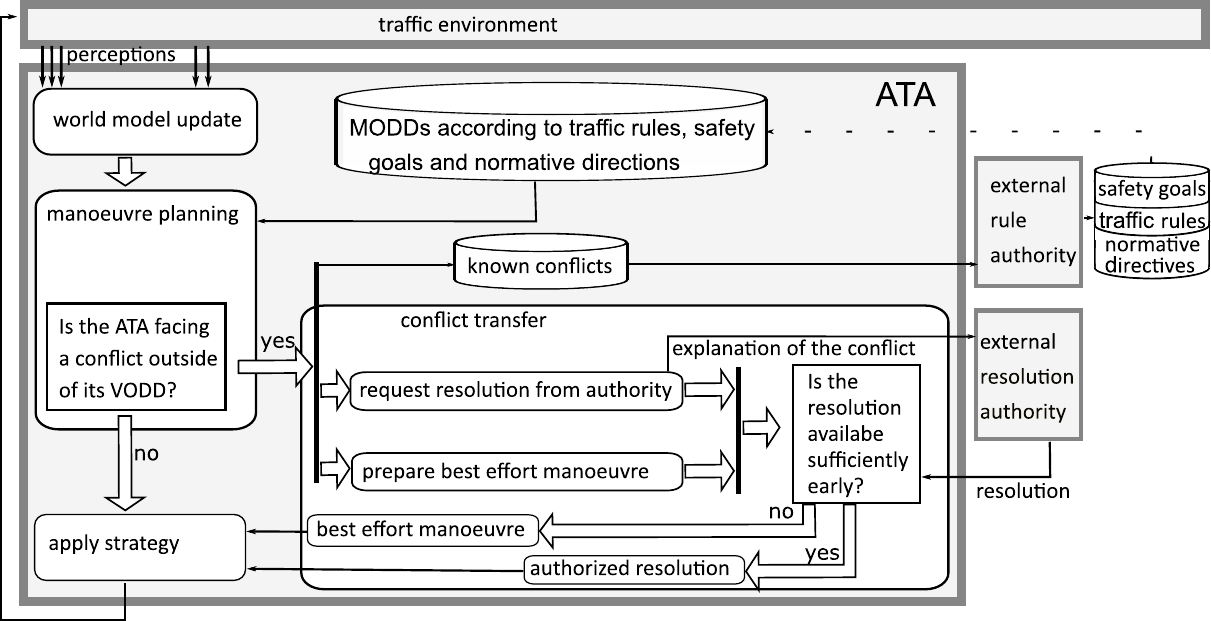}
		\caption{Conflict-sensitive architecture for value-aligned ATAs, with a focus on structured autonomy management under VODD-based scoping. The architecture integrates detection, resolution, transfer, and explanation components.}
		
	\label{fig:architecture}
\end{figure}

\autoref{fig:architecture} shows an architecture built around a sense plan act loop, extended by components for scoping and managing value-aligned autonomy.

\emph{Manoeuvre Planning and \MODDs.} The manoeuvre planner uses a database of \MODDs--contextual autonomy scopes defined at design time. They  allow the planner to delegate tasks to value-aligning autonomous components. 
Conflict detection may trigger a transfer of control, according to the exit conditions of the active \MODD.

\emph{Conflict Transfer.}  A dedicated \emph{conflict transfer component} ensures that the conflict resolution is transferred to the designated authority in a timely handover and safe manner. This includes estimating the last feasible point for receiving a resolution and preparing a fallback manoeuvre to mitigate harm if no resolution is received in time.
As part of the transfer, the transfer control component generates explanations based on the \ATA's internal conflict model (cf. \autoref{sec:expl}, \cite{Explain2025}) and tailored to the recipient prior knowledge. 
If the transfer fails, a predefined best-effort manoeuvre is initiated to mitigate harm.

\emph{Conflict Knowledge Base.} To support accountability and learning, the agent maintains a database of \emph{known conflicts}, collecting unresolved and novel conflicts encountered during operation. These also inform external rule authorities how to further evolve rules and regulations for social acceptability. 

\emph{External Authorities.} Interfaces connect the \ATA to (i) conflict-resolving au\-thorities--such as drivers, remote operators, or ethical components--to whom resolution is delegated as specified by the active \MODD, and (ii) rule authorities that issue updates to normative constraints. 
It remains an open question how best to adapt the \ATA to evolving rules. Although \MODDs are defined at design time to align with the \ATA's safety goals, traffic rules, and normative directives, modifying their entry and exit conditions may provide a mechanism for runtime adaptation. 
However, as discussed in \autoref{sec:moral} and \autoref{sec:development}, many steps in designing value-aligned autonomy are not yet automated. The development and evolution of the \MODD network thus remains a broader open challenge.

\section{Challenges and Outlook}\label{sec:challenges}
Designing conflict-sensitive autonomous traffic agents (\ATAs) involves interdependent technical, normative, and interactional challenges. While our framework provides a basis for structuring value-aligned autonomy via \MODDs, and the underlying conflict model supports reasoning about normatively significant situations, several methodological and implementation aspects remain unresolved and must be addressed to make this approach practical and robust.

\paragraph{Decision-Making Under Real-Time Constraints.}
\ATAs must select and respond to conflicts in real-time, managing uncertainties in perception, communication, and stakeholder expectations. While this is a classical concern in autonomous systems, reasoning over stakeholder values--particularly under time pressure--introduces additional complexity. Agents must be equipped with robust context-awareness, predictive capabilities, and appropriate communication infrastructure.

\paragraph{Moral Specification and Agency.}
To make morally acceptable decisions, \ATAs require explicit representations of stakeholder values and moral boundaries. Structured conflict scenarios aid in identifying value tensions, but it remains an open question how to define when an agent has failed morally. Formalising the notion of a moral agent is critical for building public trust in autonomous systems.

\paragraph{Human Oversight and Explanation.}
When conflict resolution is transferred, \ATAs should provide timely explanations that highlight the most relevant information~\cite{relevance2024,relevance2023}. These explanations must be tailored to the receiving authority--human or institutional--by identifying causally significant factors, aligning with the recipients mental model, and meeting normative and legal requirements. Designing such explanation mechanisms remains a multidisciplinary challenge spanning AI, cognitive science, legal reasoning, and human factors engineering.

\section{Summary and Conclusion}\label{sec:concl}

We presented a conflict-sensitive framework for designing value-aligned Autonomous Traffic Agents (\ATAs). By modelling value tensions as structured epistemic conflicts--arising from \goals, \beliefs, and \mykey{behaviours}--we enable systematic identification, explanation, and resolution of such tensions across the system lifecycle. This supports the informed design of agent capabilities, autonomy boundaries, and decision-making procedures aligned with stakeholder values.

We introduced Value-Aligned Operational Design Domains (\MODDs), which extend traditional ODDs by embedding explicit normative structures into auto\-nomous decision-making. \MODDs guide value-aligned behaviour by specifying prioritised goal hierarchies and contextual conditions for control transfer, enabling agents to act appropriately in both predictable and unforeseen scenarios while respecting normative and functional constraints.

Our approach addresses several key challenges: How can stakeholder values be reliably elicited, prioritised, and modelled in diverse, dynamic contexts? How can autonomy remain both adaptable and accountable? And how can agents balance social expectations with operational feasibility while ensuring transparency and explainability?\\

\textbf{Acknowledgements} {Thanks to colleagues at DLR for helpful feedback. Language editing was supported by OpenAI's ChatGPT; all content remains the author's own.}

   

\bibliographystyle{splncs04}
 \bibliography{refs}

 \end{document}